\documentclass[aps,pra,twocolumn,showpacs,superscriptaddress]{revtex4-1}

\usepackage[T1]{fontenc}
\usepackage[latin9]{inputenc}
\usepackage{amsmath}
\usepackage{amssymb}
\usepackage{ifthen}
\usepackage{times}
\usepackage{graphicx}
\usepackage{psfrag}

\def\be{\begin{equation}}
\def\ee{\end{equation}}
\def\bea{\begin{eqnarray}}
\def\eea{\end{eqnarray}}
\def\bes{\begin{subequations}}
\newcommand{\besl}[1]{\begin{subequations}\label{#1}\begin{equation}}
\def\ees{\end{subequations}}
\def\beas{\begin{subequations}\begin{eqnarray}}
\newcommand{\beasl}[1]{\begin{subequations}\label{#1}\begin{eqnarray}}
\def\eeas{\end{eqnarray}\end{subequations}}
\def\bi{\begin{itemize}}
\def\ei{\end{itemize}}

\usepackage{amssymb,amsmath}
\usepackage[usenames]{color}
\usepackage{graphicx}
\newcommand{\op}[1]{\hat{#1}}
\newcommand{\dagop}[1]{\hat{#1}^{\dagger}}
\newcommand{\bo}[1]{{\mathbf{#1}}}
\newcommand{\mc}[1]{{\mathcal{#1}}}
\newcommand{\wt}[1]{{\widetilde{#1}}}

\newcommand{\nonu}{\nonumber}
\newcommand{\etal}{~\textsl{et al.}}

\newcommand{\bra}[1]{\langle#1\vert}
\newcommand{\ket}[1]{\vert#1\rangle}

\newlength{\templength}

\newcommand{\REM}[1]{\ifthenelse{0=1}{#1}{}}

\begin{document}

%\title{A Bogoliubov method for colliding condensates using the positive-P representation}
\title{Bogoliubov dynamics of condensate collisions using the positive-P representation}

\author{P. Deuar}
\affiliation{Institute of Physics, Polish Academy of Sciences, Al. Lotnik\'ow 32/46, 02-668 Warsaw, Poland}

\author{J. Chwede\'nczuk}
\affiliation{Institute of Theoretical Physics, Physics Department, University of Warsaw, Ho\.{z}a 69, PL-00-681 Warsaw, Poland}

\author{M. Trippenbach}
\affiliation{Institute of Theoretical Physics, Physics Department, University of Warsaw, Ho\.{z}a 69, PL-00-681 Warsaw, Poland}

\author{P. Zi\'n}
\affiliation{The Andrzej So{\l }tan Institute for Nuclear Studies, Ho\.{z}a 69, PL-00-681 Warsaw, Poland}

\begin{abstract}
  We formulate the time-dependent Bogoliubov dynamics of colliding Bose-Einstein condensates
  in terms of a positive-P representation of the Bogoliubov field. We obtain stochastic evolution equations for the
  field which converge to the full Bogoliubov description as the number of realisations grows.
  The numerical effort grows linearly with the size of the computational lattice.
  We benchmark the efficiency and accuracy of our description against Wigner distribution and exact positive-P methods.
  We consider its regime of applicability, and show that it is the most efficient method in the common situation --
  when the total particle number in the system is insufficient for a truncated Wigner treatment.
\end{abstract}
\maketitle

% Powiedziec o modach co to jest.

\section{Introduction}

The collision of two Bose-Einstein condensates (BECs) -- if the relative velocity is sufficiently high -- leads to the formation of a halo of scattered atoms.
This phenomenon has been the object of numerous experimental
\cite{Kozuma99,Deng99,Chikkatur00,Maddaloni00,Band01,Steinhauer02,Vogels02,Katz02,Vogels03,Katz04,Buggle04,Katz05,Perrin07,Perrin08,Dall09,Krachmalnicoff10,Jaskula10}
and theoretical investigations
\cite{Band00,Trippenbach00,Yurovsky02,Bach02,Chwedenczuk04,Norrie05,Zin05,Katz05,Chwedenczuk06,Zin06,Norrie06,Deuar07,Drummond07,
  Chwedenczuk08,Perrin08,Molmer08,Ogren09,Deuar09,Wang10,Krachmalnicoff10}.
The atoms forming the halo could be used for precision measurements \cite{Bachor04}, interferometry
\cite{Gross10,Bouyer97,Dunningham02,Campos03,Jaskula10}, or tests of quantum
mechanics \cite{Reid09}. Condensate collisions are also related to such phenomena as
molecular dissociation
\cite{Ogren08,Mukaiyama04,Durr04,Greiner05,Poulsen01,
Kheruntsyan02,Kheruntsyan05a,Kheruntsyan06,Savage06,Savage07,Jack05,Zhao07,Tikhonenkov07,Davis08,Ogren09a,Ogren10},
atomic four-wave mixing
\cite{Trippenbach98,Deng99,Pu00,Duan00,Vogels02,Boyer08}, superradiant
scattering \cite{Inouye99,Moore99,Vardi02,Schneble03,
Schneble04, Yoshikawa04, Sadler07, Li08, Hilliard08, Buchmann10}, atomic parametric
down conversion
\cite{Campbell06,Ferris09,Fallani04,DeSarlo05,Gemelke05,Hilligsoe05,Molmer06},
and impact of a BEC on a barrier
\cite{Pasquini04,Pasquini06,Scott06,Scott07}.

Recently in a series of experimental studies \cite{Krachmalnicoff10,Jaskula10},
a quantitative analysis of the supersonic collisions of two
Bose-Einstein condensates was presented. It was based on stochastic Bogoliubov
equations for a particle field interacting via a contact potential.
In this manuscript we provide the details of that method. It relies on solving a set
of stochastic equations in a plane wave basis, rather than a diagonalization of
the Hamiltonian. We have found this approach to be more effective as it allows one to study
large scale multi-mode problems that {\it would not be possible} with direct
diagonalization. This is because in phase-space stochastic methods, such as presented in this work, the computational
requirements (memory, time) scale linearly with the number of modes or grid points.
%Also, the stochastic methods come at hand as soon as simple perturbative calculations \cite{Zin05} loose its validity.

Several stochastic methods have been used with success in the past to study the scattered atoms in these systems.
They treated the full atom - field system -- in contrast to a Bogoliubov expansion applied here --
using the truncated Wigner \cite{Norrie05,Norrie06,Deuar07,Deuar09} and the
positive-P representations \cite{Deuar07,Drummond07,Perrin08,Ogren09,Deuar09}.
However, these are not suitable for a majority of current experiments, including the recent metastable Helium condensate collisions
\cite{Perrin07,Perrin08,Chwedenczuk08,Dall09,Krachmalnicoff10,Jaskula10}.
The truncated Wigner approach is limited to the case when the total number of atoms in the system is much larger
than the number of necessary modes \cite{Sinatra02,Norrie06,Blakie08}, otherwise significant discrepancies (``truncation'') with full quantum dynamics appear.
The positive-P approach is complete, but has numerical instabilities that make it useful only for short times \cite{Deuar06,Deuar07,Deuar09},
often shorter than the duartion of the collision.

Instead of a full atom - field approach, a wide class of collisions is described accurately by a Bogoliubov description.
This approach is valid while the number of particles scattered during the collision is small in comparison with the total, a condition satisfied in most of the experiments.
The time-adaptive refinement, where the condensate wave function undergoes mean-field evolution, is sufficient to describe most collision experiments.
Moreover, contrary to a common fallacy, the Bogoliubov formulation takes into account the later Bose enhancement and stimulated scattering
into quasiparticle modes that can occur.

The drawback of the Bogoluibov method has been that accurate description of the real experimental situation  typically requires a computational grid with $10^6-10^7$ points.
A major contributing factor to this large lattice size is the need to resolve the supersonic wavelengths in the whole collision region. This large lattice renders a direct solution of the Bogoliubov-de Gennes evolution equations impossible. To avoid the diagonalization, one can introduce a phase space distribution for the Bogoliubov field. We call this approach
stochastic time-adaptive Bogoliubov (STAB). Here we use a positive-P representation of the scattered particles, which
differs from a previous well-known stochastic formulation \cite{Sinatra00}, which used a Wigner representation.
As is demonstrated below, the advantage of the present method is a much better signal-to-noise ratio in the calculations
for the most typical regimes of interest. As our positive-P based method bases on the broken-symmetry Bogoliubov description, it is applicable when
the scattered particles are well separated in momentum-space from the condensates.
This is the case for a wide range of supersonic phenomena, which apart from the condensate collisions include molecular dissociation,
superradiant scattering, and parametric down conversion, as well as supersonic
flow past barriers and other impurities \cite{Scott08,Sykes09,Paul07}.

The paper is organized as follows. Section~\ref{COLL} provides the Bogoliubov description of a BEC collision.
Section \ref{METHOD} introduces its positive-P representation, and describes the resulting stochastic evolution equations used for simulations.
In Section \ref{WIGNER} we compare the accuracy and efficiency of this positive-P Bogoliubov method (P-STAB) with the prior trunacted Wigner, positive-P and Wigner
Bogoliubov (W-STAB) methods for several characteristic BEC collision examples. We conclude with Section \ref{CONCLUSIONS}.

%%%%%%%%%%%%% BOGOLIUBOV: %%%%%%%%%%%%%%%%%%%%%%%%%%%%%%%%%%%%%%%%%%%%%%%%%%%%%%%%%%%%%%%%%%%%%%%%%%%%%%%%%%%%%%%%%%%%%%%%%%%%%%%%%%%%%%%%%%%%%%%%
\section{Colliding condensates - the Bogoliubov description}
\label{COLL}

We consider a zero-temperature, single-species bosonic gas.
As it is dilute, the interatomic interaction can be effectively reduced to a contact delta potential with strength $g$.
In the second quantization, the Hamiltonian reads
\begin{eqnarray*}
  \op{H}&=&\int\!\!d^3\bo{x}\,\dagop{\Psi}(\bo{x}) \left( -\frac{\hbar^2}{2m}\nabla^2 + V(\bo{x}) \right)\op{\Psi}(\bo{x})\nonumber\\
  &+&\frac g2\int\!\!d^3\bo{x}\,\dagop{\Psi}(\bo{x})\dagop{\Psi}(\bo{x})\op{\Psi}(\bo{x})\op{\Psi}(\bo{x}),
\end{eqnarray*}
where $m$ is the atomic mass, $g=4\pi\hbar^2a_s/m$ with $a_s$ being the s-wave scattering length and $V(\bo{x})$ is the external trapping potential.
The field operator $\hat\Psi(\bo{x})$ anihilates an atom at position $\bo{x}$ and satisfies the bosonic commutation relations.

In order to start the (half-) collision, a superposition of two counter-propagating mutually coherent atomic clouds is prepared by a Bragg pulse.
Simultaneously, the trapping potential is turned off. The two fractions start to move apart along the $z$ axis with relative speed $2v_{\rm rec}$,
twice the atomic recoil velocity. We define speed of sound using the density at the center of the initial condensate ($n_{\rm max}$),
obtaining $c_{\rm max}=\sqrt{gn_{\rm max}/m}$. In the supersonic limit, when $2v_{\rm rec}\gtrsim c_{\rm max}$, the gas is no more superfluid
and a certain portion of atoms is scattered incoherently out of the BECs, forming the \emph{halo}. The main focus of experiments and theory are the properties of the atoms in this halo.

In the time-dependent Bogoliubov approach (we use the simpler U(1) symmetry-breaking variety), the field operator is split into
\begin{equation}\label{Psi-delta}
\op{\Psi}(\bo{x},t) = \phi(\bo{x},t) + \op{\delta}(\bo{x},t),
\end{equation}
where $\phi(\bo{x},t)$ is the condensate wave function normalized to $N$ -- the number of particles. Its dynamics is governed by the Gross-Pitaevskii (GP) equation
\begin{equation}\label{GPeq}
  i\hbar\frac{d\phi(\bo{x},t)}{dt} = \left[-\frac{\hbar^2}{2m}\nabla^2 + g|\phi(\bo{x},t)|^2\right]\phi(\bo{x},t).
\end{equation}
The Bogoliubov field operator $\op{\delta}(\bo{x},t)$ describes the "noncondensed particles", and obeys the equation
\begin{eqnarray} \label{Bog}
  i \hbar \frac{\partial \op{\delta}(\bo{x},t) }{\partial t}&=&\left[ -\frac{\hbar^2}{2m}\nabla^2 + 2g|\phi(\bo{x},t)|^2\right] \op{\delta}(\bo{x},t)\nonumber\\
  &+&g\phi^2(\bo{x},t) \op{\delta}^\dagger(\bo{x},t).
\end{eqnarray}
The derivation of above equation is standard, and based on removing higher-order dependence on $\op{\delta}$ and $\dagop{\delta}$ -- equivalent to assuming that the influence of
the Bogoliubov field on itself is negligible as compared to the impact of the condensate.

The initial state of the trapped BEC is a solution of the stationary GP equation
\begin{equation}\label{GPeqSt}
  \mu \phi_0(\bo{x}) = \left[-\frac{\hbar^2}{2m}\nabla^2 + V(\bo{x})+ g|\phi_0(\bo{x})|^2\right]\phi_0(\bo{x}),
\end{equation}
with chemical potential $\mu$. The Bragg pulse transforms the condensate wave-function into
\begin{equation}\label{GPic}
  \phi(\bo{x},0) \propto \phi_0(\bo{x})\left[e^{ik_0 z}+e^{-ik_0 z}\right]/\sqrt{2},
\end{equation}
where $k_0=mv_{\rm rec}/\hbar$ is the wave-vector associated with the recoil velocity.
Neglecting quantum depletion, which is tiny in most cases, the state of the non-condensed particles is a vacuum, denoted by $|0\rangle$.

A common approach now would be to diagonalize the equation (\ref{Bog}) using a Bogoliubov transformation, and solve the obtained Bogoliubov-de Gennes equations.
However, for many systems of interest it requires $10^6-10^7$ points in space $\bo{x}$, which prohibits such diagonalization.
%Even more so if it is to be repeated at each time step to allow for the changes in $\phi(\bo{x},t)$ that occur as the collision progresses.

Instead, we develop an equivalent stochastic description of equation (\ref{Bog}) using the positive-P representation.
To obtain the dynamical equations, it is necessary to start from a Hamiltonian description.
The equation (\ref{Bog}), together with its conjugate, can be used to trace back the
effective Hamiltonian for the Bogoliubov field,
\begin{subequations}\label{Heff}
  \begin{eqnarray}
    \op{H}_{\rm eff} & = & \int d^3\bo{x}\, \dagop{\delta}(\bo{x})\left(- \frac{\hbar^2}{2m}\nabla^2 \right) \op{\delta}(\bo{x}) \label{Heffkin}\\
    &&+ 2g \int d^3\bo{x}\,|\phi(\bo{x})|^2 \dagop{\delta}(\bo{x})\op{\delta}(\bo{x})\label{Heffmf}\\
    &&+ \frac{g}{2} \int d^3\bo{x}\,\phi(\bo{x})^2 \dagop{\delta}(\bo{x})\dagop{\delta}(\bo{x}) + \text{ h.c. }\label{Heffpair}
  \end{eqnarray}
\end{subequations}
The line (\ref{Heffkin}) contains the kinetic energy of the noncondensed particles and (\ref{Heffmf}) the interaction between condensate and noncondensate particles.
Finally, (\ref{Heffpair}) governs the transfer of atomic pairs from the BEC to the $\op{\delta}$ field.

%%%%%%%%%%%%% PPB: %%%%%%%%%%%%%%%%%%%%%%%%%%%%%%%%%%%%%%%%%%%%%%%%%%%%%%%%%%%%%%%%%%%%%%%%%%%%%%%%%%%%%%%%%%%%%%%%%%%%%%%%%%%%%%%%%%%%%%%%
\section{Stochastic Time-Adaptive Bogoliubov (P-STAB) method}
\label{METHOD}

\subsection{Positive-P representation of the Bogoliubov field}
\label{PPBOG}

We employ the positive-P representation to expand the  density matrix for the uncondensed field $\op{\delta}(\bo{x},t)$ as a distribution $P$ over local coherent states at each point $\bo{x}$ in space,
\begin{subequations}\label{rep-rho}\begin{equation}
  \op{\rho}
  = \int P\left[\psi,\wt{\psi}\right]  \op{\Lambda}\left[\psi,\wt{\psi}\right] \mc{D}^2\psi\mc{D}^2\wt{\psi},
\end{equation}
where the complex fields $\psi(\bo{x})$ and $\wt{\psi}(\bo{x})$ are the amplitudes of
the local off-diagonal coherent state projectors $\op{\Lambda}$
\begin{eqnarray}\label{lambda}
  &&\hspace*{-0.5cm}\op{\Lambda}\left[\psi,\wt{\psi}\right] = \bigotimes_{\bo{x}} \op{\Lambda}_\bo{x}\left(\psi(\bo{x}),\wt{\psi}(\bo{x})\right).\\
  && =\quad {\mc N}\ e^{\int\psi(\bo{x})\,\dagop{\delta}(\bo{x})\,d\bo{x}} \ket{0}\bra{0} e^{\int\wt{\psi}(\bo{x})^*\op{\delta}(\bo{x})\,d\bo{x}}.\nonumber
\end{eqnarray}
with normalisation ${\mc N}= e^{-\int\wt{\psi}(\bo{x})^*\psi(\bo{x})\,d\bo{x} }$.
The operator $\ket{0}\bra{0}$ projects onto the vacuum state.
As the numerical computation is made on a grid, the local projectors $\op{\Lambda}_{\bo{x}}$ take on the form
\begin{eqnarray}\label{localLambda}
  \op{\Lambda}_{\bo{x}} &=& e^{\psi({\bo{x}})\,\dagop{\delta}({\bo{x}})\Delta V} \ket{0}\bra{0} e^{\wt{\psi}({\bo{x}})^*\left(\op{\delta}({\bo{x}})-\psi({\bo{x}})\right)\Delta V},\\
&=& \frac{ |\alpha\rangle_{\bo{x}}\langle\wt{\alpha}|_{\bo{x}} }
{ \langle\wt{\alpha}|_{\bo{x}} |\alpha\rangle_{\bo{x}}},
\end{eqnarray}\end{subequations}
where $\Delta V=\Delta x\cdot \Delta y\cdot \Delta z$ is the volume per grid point, $\alpha=\psi(\bo{x})\sqrt{\Delta V}$, $\wt{\alpha}=\wt{\psi}(\bo{x})\sqrt{\Delta V}$, and $|\alpha\rangle_{\bo{x}}$ is a coherent state at location $\bo{x}$ with complex amplitude $\alpha$.
We underline that the distribution $P\left[\psi,\wt{\psi}\right]$ contains complete information about the density matrix $\op{\rho}$.

Since it is non-negative and real, it can be regarded as a probability distribution of the complex valued fields $\psi(\bo{x})$ and $\wt{\psi}(\bo{x})$.
It is therefore \textit{also} equivalent to a large ensemble of samples of the fields. Consequently, the state $\op{\rho}$ is reproduced by
the set of $\psi(\bo{x})$ and $\wt{\psi}(\bo{x})$ when the number of samples $S$ tendts to infinity.
The assumption that the initial state of $\op{\delta}$ is vacuum translates into
\begin{equation}\label{ICvacuum}
  \psi({\bo{x}},0) = \wt{\psi}({\bo{x}},0) = 0.
\end{equation}

\subsection{Dynamics}
\label{PPDYN}

%This follows standard methods\cite{GardinerQN,GardinerHSM,Drummond80,Deuar02}.
The quantum evolution of the state
\begin{equation}\label{hr}
  i\hbar\frac{\partial\op{\rho}}{\partial t} = \left[\op{H}_{\rm eff} , \op{\rho}\right]
\end{equation}
is equivalent to a partial differential equation for $P$ \cite{GardinerQN,GardinerHSM,Drummond80,Deuar02}, which can be derrived using the operator identities
\begin{eqnarray}
\op{\delta}(\bo{x})\op{\Lambda} &=&  \psi(\bo{x}) \op{\Lambda}\\
\dagop{\delta}(\bo{x})\op{\Lambda} &=&  \left[\wt{\psi}(\bo{x})^*+\frac{1}{\Delta V}\frac{\partial}{\partial \psi(\bo{x})}\right] \op{\Lambda}\\
\op{\Lambda}\dagop{\delta}(\bo{x}) &=&  \wt{\psi}(\bo{x})^* \op{\Lambda}\\
\op{\Lambda}\op{\delta}(\bo{x}) &=&  \left[\psi(\bo{x})+\frac{1}{\Delta V}\frac{\partial}{\partial \wt{\psi}(\bo{x})^*}\right] \op{\Lambda}.
\end{eqnarray}
These identities are used to convert the quantum operators in $\op{H}_{\rm eff}$ inside Eq. (\ref{hr}) to partial derivatives.
The resulting equation is of a Fokker-Planck type and it is well known that it
can be rendered into a random walk of the samples of $\psi(\bo{x})$ and $\wt{\psi}(\bo{x})$ -- the Langevin equations -- which in the Ito representation read
\begin{widetext}
  \beasl{STABeq}
  i\hbar\frac{d\psi({\bo{x}},t)}{dt} &=& \left\{ -\frac{\hbar^2}{2m}\nabla^2+ 2g|\phi({\bo{x}},t)|^2\right\}\psi({\bo{x}},t)
  +g\,\phi({\bo{x}},t)^2\wt{\psi}({\bo{x}},t)^* + \sqrt{i\hbar g}\,\phi({\bo{x}},t) \xi({\bo{x}},t),\label{STABpsi}\\
  i\hbar\frac{d\wt{\psi}({\bo{x}},t)}{dt} &=& \left\{-\frac{\hbar^2}{2m}\nabla^2 + 2g|\phi({\bo{x}},t)|^2\right\}\wt{\psi}({\bo{x}},t)
  +g\,\phi({\bo{x}},t)^2\psi({\bo{x}},t)^* + \sqrt{i\hbar g}\,\phi({\bo{x}},t) \wt{\xi}({\bo{x}},t).\label{STABpsit}
  \eeas
\end{widetext}
Here $\xi({\bo{x}},t)$ and $\wt{\xi}({\bo{x}},t)$ are delta-correlated, independent, real gaussian stochastic noise fields with variances
\begin{eqnarray}
  &&\langle \xi({\bo{x}},t)\wt{\xi}({\bo{x}}',t') \rangle =0\label{noise}\\
  &&\hspace*{-0.5cm}\langle \xi({\bo{x}},t)\xi({\bo{x}}',t') \rangle
  = \langle \wt{\xi}({\bo{x}},t)\wt{\xi}({\bo{x}}',t') \rangle
  =\delta^{(3)}({\bo{x}}-{\bo{x}}')\delta(t-t').\nonu
\end{eqnarray}
and zero mean. Numerically, $\xi$ and $\wt{\xi}$ are usually approximated by real gaussian random variables of variance $1/(\Delta t\Delta V)$ that are independent at each point
at the computational lattice, and at each time step of length $\Delta t$.

%These stochastic equations stricly reproduce the full quantum dynamics described by $\op{H}_{\rm eff}$ as the number of samples of the fields tends to infinity.
%For finite sample sizes, the difference is random and can be easily approximated from the samples at hand using the central limit theorem.
%The details of the stochastic phase-space methods are best described elsewhere \cite{GardinerQN,GardinerHSM}.
One important feature of the equations (\ref{STABeq}) is that, similarly to Eq.(\ref{Bog}),  they are \emph{linear} in $\psi$ and $\wt{\psi}$.
This way, the nonlinear instabilities are absent, together with boundary term systematics \cite{Gilchrist97,Deuar02}
and finite-simulation-time issues \cite{Deuar06} that may occur in direct positive-P treatments of the full boson field $\op{\Psi}$.

\subsection{Observables}
\label{OBS}
Expectation value of any normal-ordered observable are evaluated using the positive-P representation by substituting $\dagop{\delta}\to\wt{\psi}^*$ and $\op{\delta}\to\psi$
and calculating a stochastic average \cite{Deuar02}, denoted as $\langle\cdot\rangle_{\rm st}$,
\begin{equation}\label{norm-order}
  \Big\langle\prod_j\dagop{\delta}(\bo{x}_j)\prod_k\op{\delta}({\bo{x}}_k)\Big\rangle
  = \lim_{S\to\infty}\Big\langle\prod_j \wt{\psi}({\bo{x}}_j)^* \prod_k \psi({\bo{x}}_k)\Big\rangle_{\rm st}.
\end{equation}
Note that since Eq.(\ref{Bog}) is linear, and the initial state is vacuum, then at all times $t$
\begin{equation}\label{statedelta}
\langle 0| \op{\delta}(\bo{x},t) |0 \rangle = 0.
\end{equation}
As an example, the one-particle density matrix is given by
\begin{eqnarray*}
  \rho_1(\bo{x},\bo{x}',t)&=&\left\langle\dagop{\Psi}(\bo{x},t)\op{\Psi}(\bo{x}',t)\right\rangle \\
  &=& \phi(\bo{x},t)^*\phi(\bo{x}',t) + \langle \dagop{\delta}(\bo{x},t) \op{\delta}(\bo{x}',t)\rangle\\
&& +\phi(\bo{x},t)^*\langle\op{\delta}(\bo{x}',t)\rangle + \phi(\bo{x}',t)\langle \dagop{\delta}(\bo{x},t)\rangle \\
  &=&\phi(\bo{x},t)^*\phi(\bo{x}',t) + \langle \dagop{\delta}(\bo{x},t) \op{\delta}(\bo{x}',t)\rangle \\
  &=&\phi(\bo{x},t)^*\phi(\bo{x}',t) + \langle\wt{\psi}(\bo{x},t)^*\psi(\bo{x}',t)\rangle_{\rm st}
\end{eqnarray*}
%% \beas
%% &\rho_1(\bo{x},\bo{x}',t)&= \left\langle\dagop{\Psi}(\bo{x},t)\op{\Psi}(\bo{x}',t)\right\rangle \\
%% &=& \left\langle \left(\phi(\bo{x},t)^*+\dagop{\delta}(\bo{x},t)\right)\left(\phi(\bo{x}',t)+\op{\delta}(\bo{x}',t)\right)\right\rangle\nonu\\
%% &=& \phi(\bo{x},t)^*\phi(\bo{x}',t) + \langle\dagop{\delta}(\bo{x},t)\op{\delta}(\bo{x}',t)\rangle\nonu\\
%% && + \phi(\bo{x},t)^*\langle\op{\delta}(\bo{x}',t)\rangle + \phi(\bo{x}',t)\langle\dagop{\delta}(\bo{x},t)\rangle \\
%% &=& \phi(\bo{x},t)^*\phi(\bo{x}',t) + \langle\dagop{\delta}(\bo{x},t)\op{\delta}(\bo{x}',t)\rangle\label{approx-1}\\
%% &=& \phi(\bo{x},t)^*\phi(\bo{x}',t) + \textrm{Re}\langle\wt{\psi}(\bo{x},t)^*\psi(\bo{x}',t)\rangle_{st}.\label{obs-rho1-primary}
%% \eeas
where we used  (\ref{statedelta}) on the second-last line.
The number of non-condensed atoms is
\begin{equation}
\delta N = \int \langle\dagop{\delta}(\bo{x})\op{\delta}(\bo{x})\rangle \, d^3\bo{x}
= \int \langle\wt{\psi}(\bo{x})^*\psi(\bo{x})\rangle\rangle_{\rm st}\,d^3\bo{x}.
\end{equation}

For a general observable $\op{F}$, the best estimate of its expectation value $\langle\op{F}\rangle$ is given by the mean of its corresponding estimator $f(\psi,\wt{\psi},\phi)$
$$
\bar{F} = \langle f \rangle_{\rm st}.
$$
The uncertainty in this mean is best estimated via the variance of a set of subensemble means: We divide the $S$ realizations into $n$ bins of equal size $s$ (so $S=sn$), and
the $j$th subensemble ($j=1,\dots,n$) gives a subensemble mean $\bar{F}_j = \langle f \rangle_{\rm st,j}$. Due to the central limit theorem,
these subensemble means are approximately normal distributed (which is not necessarily the case for the estimators from individual realizations).
As a result, the uncertainty in the final mean ($\bar{F}=\frac{1}{n}\sum_{j=1}^n \bar{F}_j$ also) is well estimated by
\begin{equation}\label{uncert}
\Delta F = \sqrt{\frac{\text{var}\left[\bar{F}_j\right]}{n-1}}.
\end{equation}

\subsection{Orthogonality and applicability}
\label{ORTH}

It is well known that the U(1) symmetry breaking Bogoliubov method reveals some problems at longer evolution times.
These are related to an incomplete treatment of the the phase spreading of the condensate \cite{Lewenstein96}.
As the approach does not preserve the orthogonality of the non-condensed field $\op{\delta}(\bo{x})$ to the condensate mode \cite{Castin98},
the part of $\op{\delta}$ that accumulates atop the condensate could just as well be considered to still be part of the BEC, and discounted from the number of
scattered particles.

For this reason, the results of the above method should be treated with caution when modes having significant overlap with the condensate are relevant. In practice, such modes
lie in parts of k-space close to the condensate clouds. Fortunately, the bulk of the halo is well separated from the condensates and remains unaffected.

More generally, supersonicity always leads to orthogonality between scattered and condensed atoms because the condensate mode function
contains no plane-wave components above the speed of sound. This allows the use of the method presented here for collisions of BECs, molecular dissociation,
superradiant scattering, parametric down conversion, or flow past barriers and other impurities.
\begin{figure}
  \begin{centering}
    \resizebox{\columnwidth}{!}{
      \includegraphics{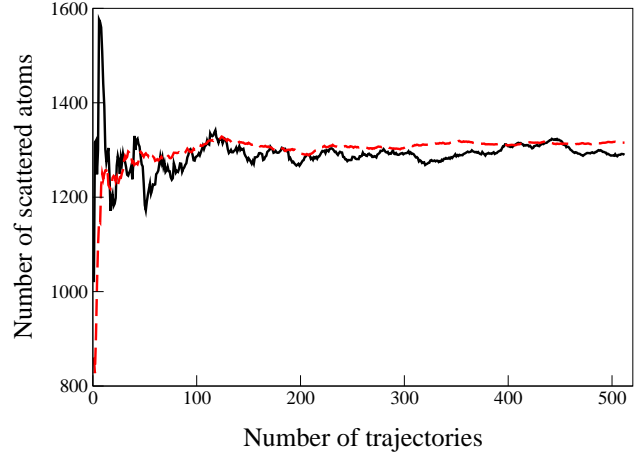}
    }
  \end{centering}
  \vspace*{-10pt}
  \caption{\label{Fig-nscatm}
The convergence of observable estimates  in the two Bogoliubov methods as the number of trajectories is increased.
The system is the ${}^4$He collision of \cite{Krachmalnicoff10}. The quantity shown is the total number of atoms in the halo at $t=120\mu s$, well after the end of the collision.
Narrow k-space regions containing the condensates were excluded from the atom sum.
Black solid line: Wigner Bogoliubov calculation (W-STAB), Red dashed: Positive-P Bogoliubov (P-STAB).
  }
  \vspace*{-15pt}
\end{figure}

%%%%%%%%%%%%% WIGNER %%%%%%%%%%%%%%%%%%%%%%%%%%%%%%%%%%%%%%%%%%%%%%%%%%%%%%%%%%%%%%%%%%%%%%%%%%%%%%%%%%%%%%%%%%%%%%%%%%%%%%%%%%%%%%%%%%%%%%%%
\section{Relationship with comparable methods}
\label{WIGNER}

Stochastic evolution equations have been previously derived for Bogoliubov descriptions of cold atoms systems by Sinatra\etal\cite{Sinatra00} using the Wigner representation.
An immediate question is how the positive-P based method presented here compares. We expect that the positive-P method will tend to be inherently less ``noisy'' initially
due the lack of starting noise which is necessary to represent the vacuum in the Wigner treatment.
It is also instructive to compare performance and accuracy with the two other stochastic methods used previously (positive-P and truncated Wigner) which treat the
whole atom field $\op{\Psi}$ as one unit without using the Bogoliubov approximation. In this section we will
benchmark these four simulation methods.

\subsection{Wigner STAB}
Representing the U(1) symmetry breaking description of Sec.~\ref{COLL} using the Wigner representation we obtain the following stochastic description of the field $\op{\delta}(\bo{x})$.
There is only one complex field $\psi_w(\bo{x})$, with the initial vacuum described by a random initial condition that places half a virtual particle into each mode
\begin{equation}\label{ICvacuumBW}
  \psi_w({\bo{x}},0) = \frac{\sqrt{\Delta t}}{2}\left[\xi(\bo{x},0)+i\wt{\xi}(\bo{x},0)\right].
\end{equation}
(The noises $\xi$ and $\wt{\xi}$ are as defined by (\ref{noise})\,).
The subsequent evolution contains no noise and is
\begin{eqnarray}
  \label{STABeqBW}
  i\hbar\frac{d\psi_w({\bo{x}},t)}{dt} &=& \left\{ -\frac{\hbar^2}{2m}\nabla^2+ 2g|\phi({\bo{x}},t)|^2\right\}\psi_w({\bo{x}},t)\nonu\\
  &&+g\,\phi({\bo{x}},t)^2\psi_w({\bo{x}},t)^*.
\end{eqnarray}
Observable calculations differ somewhat because the half-particle occupation of the initial modes must be corrected for. For example,
\begin{eqnarray}
  \lefteqn{\rho_1(\bo{x},\bo{x}',t)=} \\
&&\phi(\bo{x},t)^*\phi(\bo{x}',t) + \langle\psi_w(\bo{x},t)^*\psi_w(\bo{x}',t)\rangle_{\rm st} -\frac{1}{2}\delta(\bo{x}-\bo{x}').\nonu
\end{eqnarray}
This is the symmetry-breaking analogue of the more involved number-conserving description of Sinatra\etal\cite{Sinatra00}, and shares the same noise properties. However
the same orthogonality caveats (Sec.~\ref{ORTH}) apply as for the P-STAB method derived in this paper.
\begin{figure}
%\fbox{
  \begin{picture}(7.9,11.8)
    \put(-0.2,6){\includegraphics[width=8.0cm]{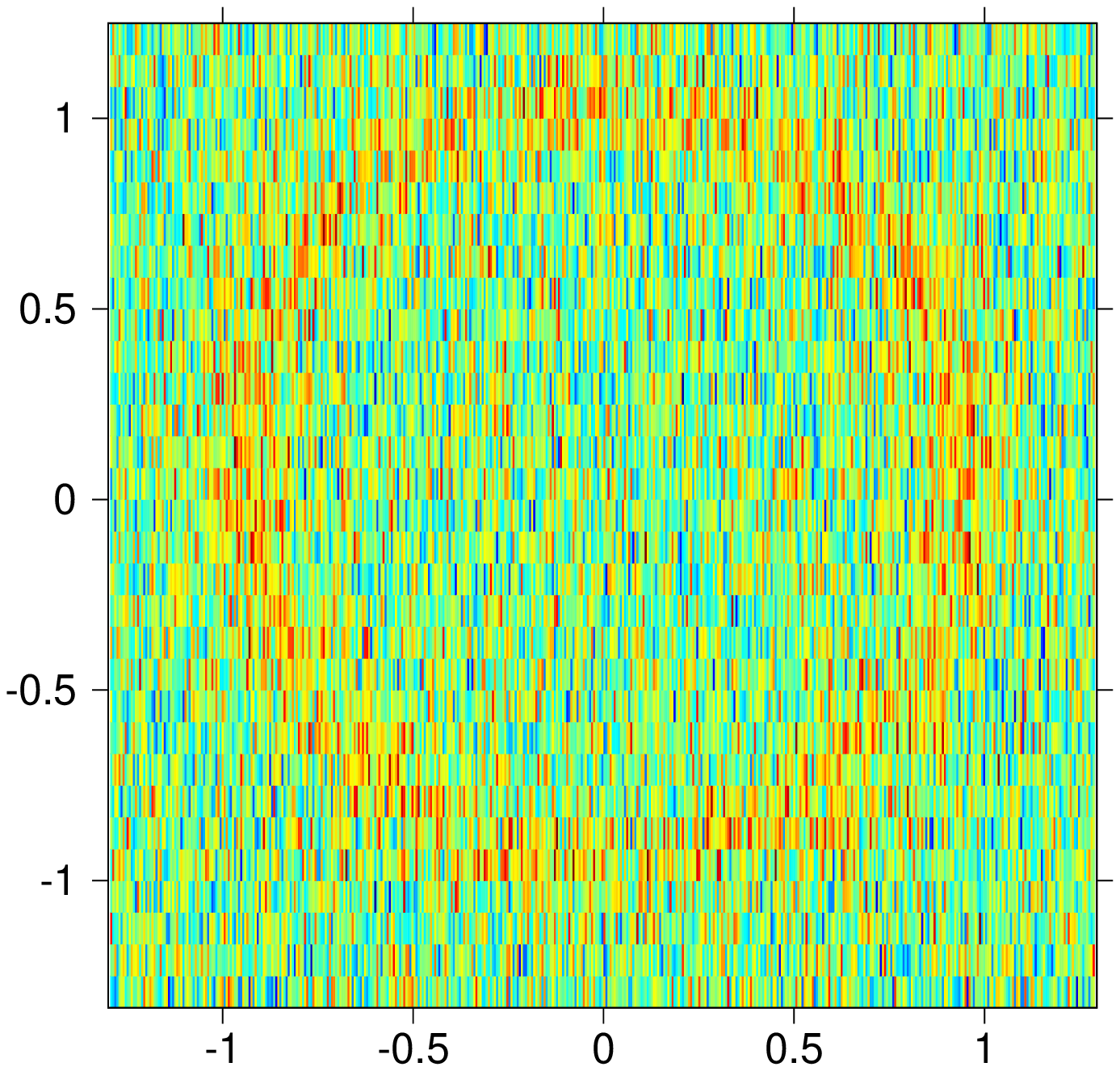}}
    \put(1.6,11.1){\bf(a)}
    \put(3.4,6.0){$k_x/k_0$}
    \put(0.5,8.7){\rotatebox{90}{$k_y/k_0$}}
    \put(-0.2,0){\includegraphics[width=8.0cm]{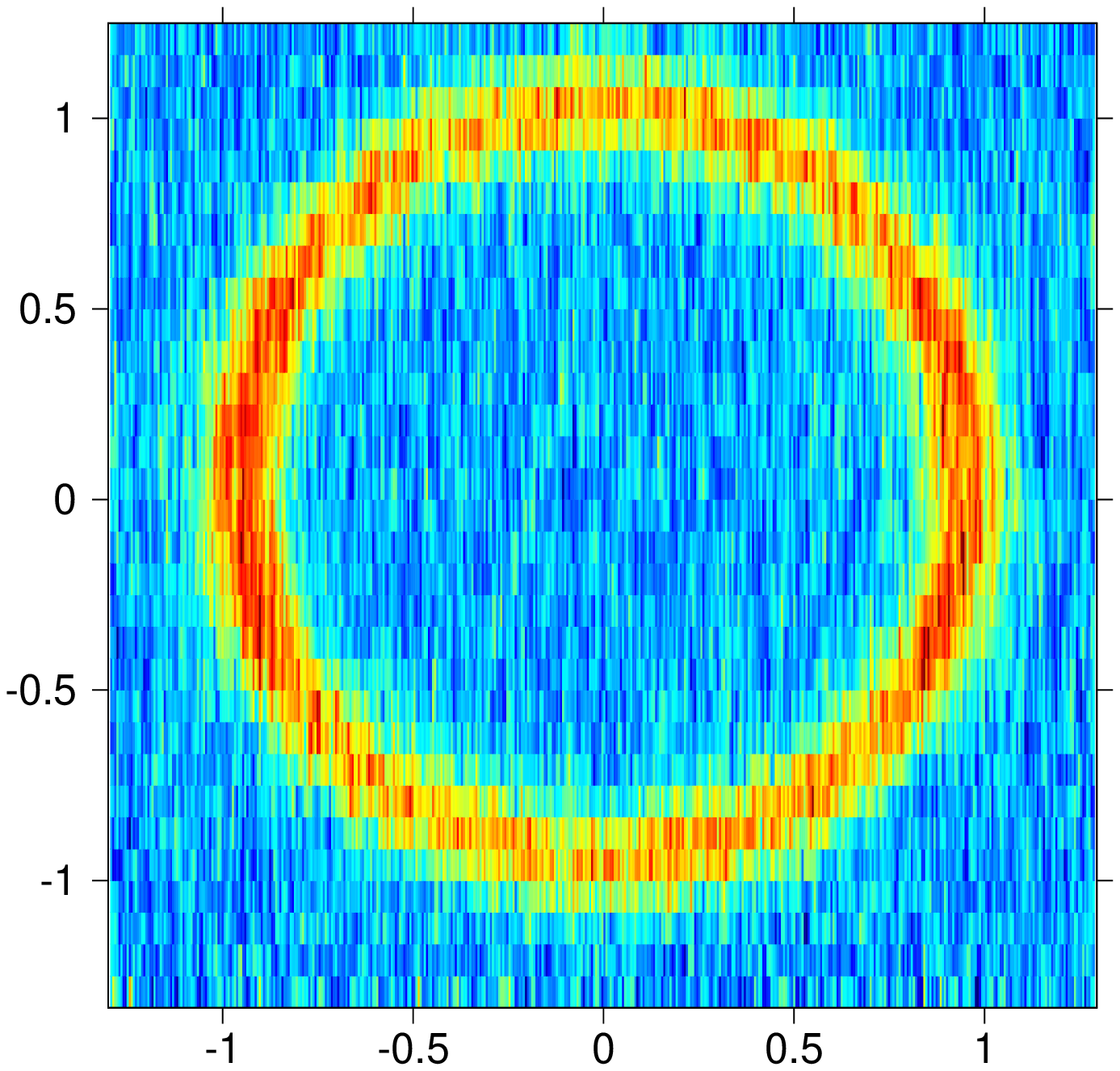}}
    \put(1.6,5.1){\bf(b)}
    \put(3.4,0.0){$k_x/k_0$}
    \put(0.5,2.7){\rotatebox{90}{$k_y/k_0$}}
  \end{picture}
%}
  \vspace*{-0.3cm}
  \caption{\label{Fig-slices}
Slices of the halo density on the plane $k_z=0$, perpendicular to the collision direction, for the ${}^4$He collision of \cite{Krachmalnicoff10}.
$t=48\mu s$, right at the end of the collision.
Both results are from ensembles of 224 realizations.
(a): Wigner Bogololiubov, (b): Positive-P Bogoluiubov.
  }
  \vspace*{-15pt}
\end{figure}

\subsection{Full-field methods}
For some parameters, another good alternative is to use the truncated Wigner representation to simulate the complete boson field directly,
as was done by Norrie \etal\ \cite{Norrie05,Norrie06}. This has the advantage of being applicable beyond the undepleted source approximation. However, the total number
of particles should be significantly larger than the number of modes (for correctness \cite{Norrie06,Deuar07}).
This approach requires the truncation of some high-order terms in the partial differential equation for the
resulting phase space distribution $P$, leading to the name ``truncated'' Wigner representation. Here there is
one complex field $\psi_W(\bo{x})$ (no separate condensate field $\phi(\bo{x},t)$) with the initial state
\begin{equation}\label{ICvacuumW}
  \psi_W({\bo{x}},0) = \phi(\bo{x},0)+\frac{\sqrt{\Delta t}}{2}\left[\xi(\bo{x},0)+i\wt{\xi}(\bo{x},0)\right].
\end{equation}
The subsequent evolution contains no noise and is
\begin{eqnarray}
  \label{STABeqW}
  i\hbar\frac{d\psi_W({\bo{x}},t)}{dt} &=& \left\{ -\frac{\hbar^2}{2m}\nabla^2+ g|\psi_W({\bo{x}},t)|^2\right\}\psi_W({\bo{x}},t).\qquad
\end{eqnarray}
The one-particle density matrix is given by
\begin{equation}
  \rho_1(\bo{x},\bo{x}',t)=
\langle\psi_W(\bo{x},t)^*\psi_W(\bo{x}',t)\rangle_{\rm st} -\frac{1}{2}\delta(\bo{x}-\bo{x}').
\end{equation}

Finally, a direct treatment of the full field using the positive-P representation has been used\cite{Deuar07,Drummond07,Perrin08,Ogren09,Deuar09}.
Here there are two
complex fields $\psi_p(\bo{x})$ and $\wt{\psi}_p(\bo{x})$ with the initial state
\begin{equation}\label{ICvacuumPP}
  \psi_p({\bo{x}},0) = \wt{\psi}_p(\bo{x},0) = \phi(\bo{x},0).
\end{equation}
The evolution is
\begin{eqnarray}
  \label{STABeqPP}
  i\hbar\frac{d\psi_p({\bo{x}},t)}{dt} &=& \left\{ -\frac{\hbar^2}{2m}\nabla^2+ g\wt{\psi}_p(\bo{x},t)^*\psi_p(\bo{x},t)\right.\nonu\\
&&\left.+\sqrt{i\hbar g}\,\xi(\bo{x},t)\right\}\psi_p({\bo{x}},t)\\
  i\hbar\frac{d\wt{\psi}_p({\bo{x}},t)}{dt} &=& \left\{ -\frac{\hbar^2}{2m}\nabla^2+ g\psi_p(\bo{x},t)^*\wt{\psi}_p(\bo{x},t)\right.\nonu\\
&&\left.+\sqrt{i\hbar g}\,\wt{\xi}(\bo{x},t)\right\}\wt{\psi}_p({\bo{x}},t).
\end{eqnarray}
The one-particle density matrix is calculated with
\begin{equation}
  \rho_1(\bo{x},\bo{x}',t)=
\langle\wt{\psi}_p(\bo{x},t)^*\psi_p(\bo{x}',t)\rangle_{\rm st}.
\end{equation}

\subsection{Efficiency measures}

\begin{figure}
  \begin{centering}
    \resizebox{\columnwidth}{!}{
      \includegraphics{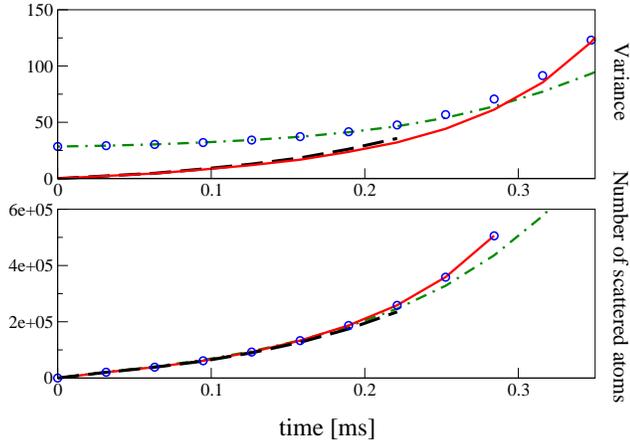}
    }
  \end{centering}
  \vspace*{-10pt}
  \caption{\label{Fig-norr}
    ${}^{23}$Na BEC collision with the same parameters as used in \cite{Norrie05}. $N=6\times10^6$
    Upper panel: Variances of local atom density estimators in the slice at $k_z=0$ obtained for various methods, as for use in (\ref{uncert}) -- see text.
    An average value over all $k_x$ and $k_y$ locations in the slice is shown.
    Lower panel: Number of scattered atoms in the halo.
    Solid red: Positive-P Bogoliubov simulation as described in this paper;
    Blue circles: Wigner Bogoliubov simulation;
    Dot-dash green: truncated Wigner simulation;
    Dashed black: positive-P simulation of the full field.
    The Bogoliubov simulations were stopped when the depletion reached 10\%.
  }
  \vspace*{-15pt}
\end{figure}

When considering the halo, the most pertinent observables have been the total number of particles, the density distribution in k-space, and density correlations
between specified regions in the halo. Accuracy in the latter two kinds of observables hinge on a good signal-to-noise ratio of the local density in k-space.
The uncertainty of the final estimates is given by (\ref{uncert}), a function of the ratio between variance of the estimator and the number of realizations $S\propto n$.
So, other things being equal, the computational effort required to achieve a set accuracy will scale as that variance.
Accordingly, in Figs.~\ref{Fig-na} and~\ref{Fig-norr} (upper panels) we will show how the variance of the estimators of halo density in k-space compare between methods as a function of time.
In Figs.~\ref{Fig-nscatm} and~\ref{Fig-slices} we directly show the noise that is seen with the full-field positive-P and Wigner Bogoliubov treatments.

\subsection{The low and moderate particle number case}

Let us first consider the common case when the total number of atoms in the halo is quite low -- so low that the number of halo atoms per mode is much less than one.
Here we expect the initial noise in the Wigner methods (\ref{ICvacuumBW}) or (\ref{ICvacuumW}) to be a severe problem, since the initial atom number variance there is
$1/2$ per mode, regardless of how many true atoms are present.

The first plots (Figs.~\ref{Fig-nscatm} and~\ref{Fig-slices}) are from Wigner and positive-P Bogoliubov simulations using the experimental parameters of \cite{Krachmalnicoff10},
which described the collision of a BEC of metastable ${}^4{\rm He}^*$ atoms. They show the amount of noisyness in observables after the end of the collision.
In this case, no bosonic enhacement of the scattering process occured, thus the total number of atoms in the halo was quite low ($\approx$1300), while the number of modes
was $2.95\times10^6$.

The next figure,~\ref{Fig-na}, shows the halo density variance and the total number of scattered atoms
in the collision of a BEC of 150 000  ${}^{23}$Na atoms. This case was considered in several previous works\cite{Deuar07,Deuar09,Ogren09}.
Here the halo reached $1.1\times10^4$ atoms with $1.08\times10^6$ modes).

We see that the noise in the Wigner calculations is severe in these cases, as compared to the positive-P methods.
Although the noise in the P-STAB calculation grows with time, it never surpasses the
level of the Wigner methods for the timescales shown. The variance in both Wigner methods is identical.

The lower panel of Fig.~\ref{Fig-na} shows the accuracy of the methods. Both Bogoliubov methods agree perfectly with each other, and with the exact calculation that uses the
positive-P representation of the full field (for as long as it lasts).  The truncated Wigner displays a false growth of the number of particles in the halo.
This is due to known spurious scattering by virtual particles when the momentum cutoff is this large, as described in \cite{Sinatra02,Deuar07,Deuar09}.
For this simulations, the number of spatial modes is much larger than the number of true particles (150 000).

\begin{figure}
  \begin{centering}
    \resizebox{\columnwidth}{!}{
      \includegraphics{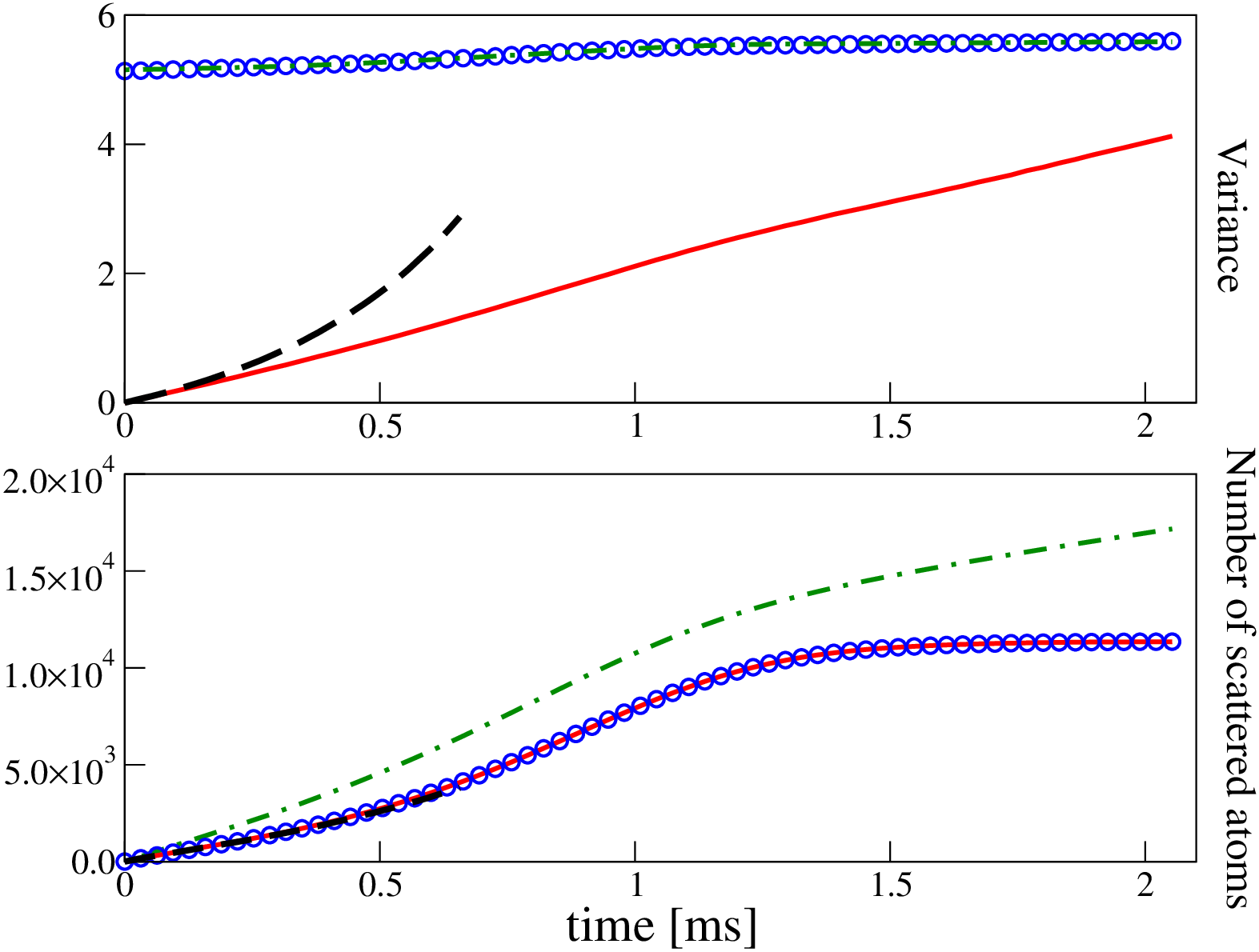}
    }
  \end{centering}
  \vspace*{-10pt}
  \caption{\label{Fig-na}
    ${}^{23}$Na BEC collision as in \cite{Deuar07,Deuar09,Ogren09}. $N=1.5\times10^5$
    Upper panel: Variances of local atom density estimators in the slice at $k_z=0$ obtained for various methods, as for use in (\ref{uncert}) -- see text.
    An average value over all $k_x$ and $k_y$ locations in the slice is shown.
    Lower panel: Number of scattered atoms in the halo.
    Solid red: Positive-P Bogoliubov simulation as described in this paper;
    Blue circles: Wigner Bogoliubov simulation;
    Dot-dash green: truncated Wigner simulation;
    Dashed black: positive-P simulation of the full field.
  }
  \vspace*{-15pt}
\end{figure}

\subsection{The high particle number case}

A different situation is presented in (Fig.~\ref{Fig-na}), where we used the parameters from \cite{Norrie05}, where $6\times10^6$ atoms of ${}^{23}$Na participated in the collision.
There were $3.14\times10^6$ spatial modes.
As final depletion of the condensate is large (about 40\%), and the Bogoliubov calculation
must was stopped at $t\approx280\mu$s, when depletion was $10\%$. Indeed, in the lower panel of
Fig~\ref{Fig-na}, one sees a difference beginning to appear between the two simulations at this time. In comparison,
significant dynamics lasts until $\approx1000\mu$s (not shown).

The noise performance of the P-STAB method is superior here only for $t\lesssim 300\mu$s.
However, this still matches the entire period when the Bogoliubov description is accurate.

\section{Conclusions}
\label{CONCLUSIONS}

We have developed the above positive-P Bogoliubov stochastic simulation method for use with cold atom gases and benchmarked it with existing approaches.
As with other phase-spae methods, it lends itself to simulation of quite general systems, as the calculation is carried out on a simple rectangular grid in x / k space, and individual realizations are
run independently of each other. The computational complexity involved scales linarly with the size of the computational lattice used, allowing for up to $\approx10^7$
points in the lattice on a common workstation.

The method is applicable for a wide range of supersonic phenomena, its main limitations being (1) that the bulk of scattered atoms are well separated from the
condensates in momentum space, and (2) that the depletion of the original condensates can be neglected. The condensate wave function is, however, free to evolve in time.
We note particularly that the method handles both spontaneous and stimulated scattering.

The positive-P Bogoliubov method is superior in efficiency to the Wigner representation in almost all cases that we have seen where a U(1) symmetry breaking Bogoliubov method can still be applied.
However, one can imagine some long time situations where the Wigner simulation wins since, other things being equal,
the variance in the positive-P approach grows approximately linearly with time, while the variance in the Wigner method stays approximately constant around its initial, large, value
(These trends are seen in the top panel of Fig.~\ref{Fig-na} ).
For situations where the overlap between the scattered and condensate field is non-negligible the number-conserving Wigner method\cite{Sinatra00} can be used instead. For situations
with large condensate depletion, there remain the truncated Wigner or positive-P treatments of the full boson field.

A more robust positive-P formulation that explicitly imposes orthogonality between condensate and quasiparticle modes as in the number-conserving Bogoliubov treatment\cite{Castin98}
is under development and will be presented in a forthcoming work.

\begin{acknowledgments}
  It is a pleasure to thank
Karen Kheruntsyan,
Chris Westbrook,
Denis Boiron,
Nick Proukakis,
Alice Sinatra,
and Brian Dalton for valuable discussions on these matters. PD acknowledges support by the EU contract PERG06-GA-2009-256291 and Polish Government Research Funds for the years 2010-2013; MT and PZ acknowledge support of Polish Government Research Grants for 2007-2011,; JC J. C. was supported by Foundation for Polish Science International TEAM Programme co-financed by the EU European Regional Development Fund.
\end{acknowledgments}

%%%%%%%%%%%%% BIB %%%%%%%%%%%%%%%%%%%%%%%%%%%%%%%%%%%%%%%%%%%%%%%%%%%%%%%%%%%%%%%%%%%%%%%%%%%%%%%%%%%%%%%%%%%%%%%%%%%%%%%%%%%%%%%%%%%%%%%%
\newcommand{\PRL}[1]{Phys. Rev. Lett.~\textbf{#1}}
\newcommand{\PRA}[1]{Phys. Rev.~A~\textbf{#1}}

\end{document}